\documentclass[]{aa}
\usepackage[varg]{txfonts}
\usepackage{graphicx}
\usepackage{subfig}
\usepackage[pscoord]{eso-pic}
\bibpunct{(}{)}{;}{a}{}{,} % to follow the A&A style

\title{Surface flux transport simulations: Effect of inflows toward active regions and random velocities on the evolution of the Sun's large-scale magnetic field}
\author{D. Martin-Belda\inst{\ref{inst1}} \inst{\ref{inst2}}
\and R. H. Cameron\inst{\ref{inst1}}
}

\institute{Max-Planck-Institut für Sonnensystemforschung, Justus-von-Liebig-Weg 3, 37077 Göttingen, Germany\label{inst1} \and
Institut für Astrophysik, Georg-August-Universität Göttingen, 37077 Göttingen, Germany\label{inst2}
}

\date{Received / Accepted}


\abstract 
{}
{We aim to determine the effect of converging flows on the evolution of a bipolar magnetic region (BMR), and to investigate the role of these inflows in the generation of poloidal flux. We also discuss whether the flux dispersal due to turbulent flows can be described as a diffusion process.}
{We developed a simple surface flux transport model based on point-like magnetic concentrations. We tracked the tilt angle, the magnetic flux and the axial dipole moment of a BMR in simulations with and without inflows and compared the results. To test the diffusion approximation, simulations of random walk dispersal of magnetic features were compared against the predictions of the diffusion treatment.}
{We confirm the validity of the diffusion approximation to describe flux dispersal on large scales. We find that the inflows enhance flux cancellation, but at the same time affect the latitudinal separation of the polarities of the bipolar region. In most cases the latitudinal separation is limited by the inflows, resulting in a reduction of the axial dipole moment of the BMR. However, when the initial tilt angle of the BMR is small, the inflows produce an increase in latitudinal separation that leads to an increase in the axial dipole moment in spite of the enhanced flux destruction. This can give rise to a tilt of the BMR even when the BMR was originally aligned parallel to the equator.}
{}

\begin{document}
\titlerunning{Inflows and random velocities}
\authorrunning{D. Martin-Belda \& R. H. Cameron}
\maketitle

\section{Introduction}

Surface flux transport (SFT) simulations have been used with considerable success to describe the evolution of the large-scale photospheric magnetic field \citep[see, e.g.][]{devore1984meridional,wang1989evolution, mackay2002single,mackay2002full,baumann2004parameter}. These models are based on the asumption that the field at the surface is nearly radial \citep{solanki1993smallscale,mpillet1997active}, and thus its evolution can be described by the radial component of the MHD induction equation. The scalar quantity $B_r$ is advected by the large-scale flows (differential rotation and meridional flow) and the variable patterns of convection. The latter have the effect of dispersing the magnetic field, and have commonly been modeled as a Fickian diffusion process \citep{leighton1964transport}, although some authors prefer a less parametrized treatment of the turbulent dispersal. In \cite{schrijver2001simulations}, an SFT model based on discrete flux concentrations is used to simulate the evolution of the surface field. Hathaway (2010) uses an observation-based, time-evolving spectrum of spherical harmonics to produce random patterns of turbulent flows that advect magnetic concentrations. This approach recovers some of the observed characteristics of the evolution of the photospheric field, such as the accumulation of flux in the network and the dispersal on multiple scales. One of the questions we want to address in this work is whether the effects of the turbulent dispersal on the large-scale, long-term evolution of the surface field are appropriately captured by the diffusion approximation.

A second question concerns the systematic tilt of emerged bipolar magnetic regions (BMRs), which plays a central role in the Babcock-Leighton dynamo mechanism. The leading polarity of a BMR tends to emerge at lower latitudes than the trailing polarity (Joy's law), and the latter is opposite to the polarity of the polar field at the preceding activity minimum (Hale's law). The tilt angle is thought to be caused by the action of the Coriolis force on rising flux ropes \citep[see, e.g.,][]{fan2009lrsp}, and provides a mechanism for generating poloidal field from toroidal field \citep{charbonneau2010lrsp}. The latitudinal separation of the polarity patches favors the cross-equatorial transport of leading polarity, which leads to the cancellation and eventual reversal of the polar fields.
\begin{figure*}[t]
  \centering
  \subfloat[\label{fig:infexp1}]{%
    \includegraphics[width=7.4cm]{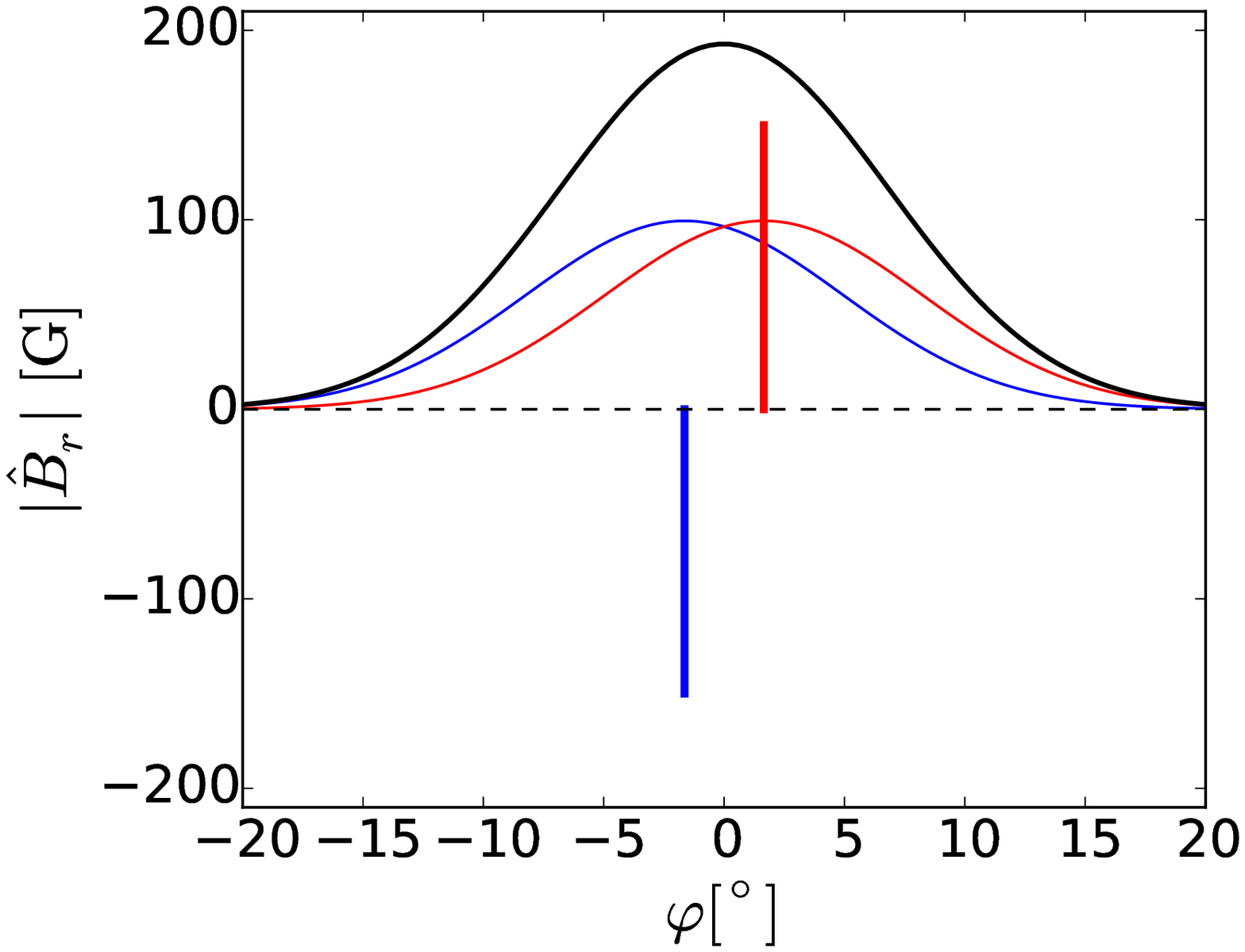}
  }\hspace{1cm}
  \subfloat[\label{fig:infexp2}]{%
    \includegraphics[width=8.6cm]{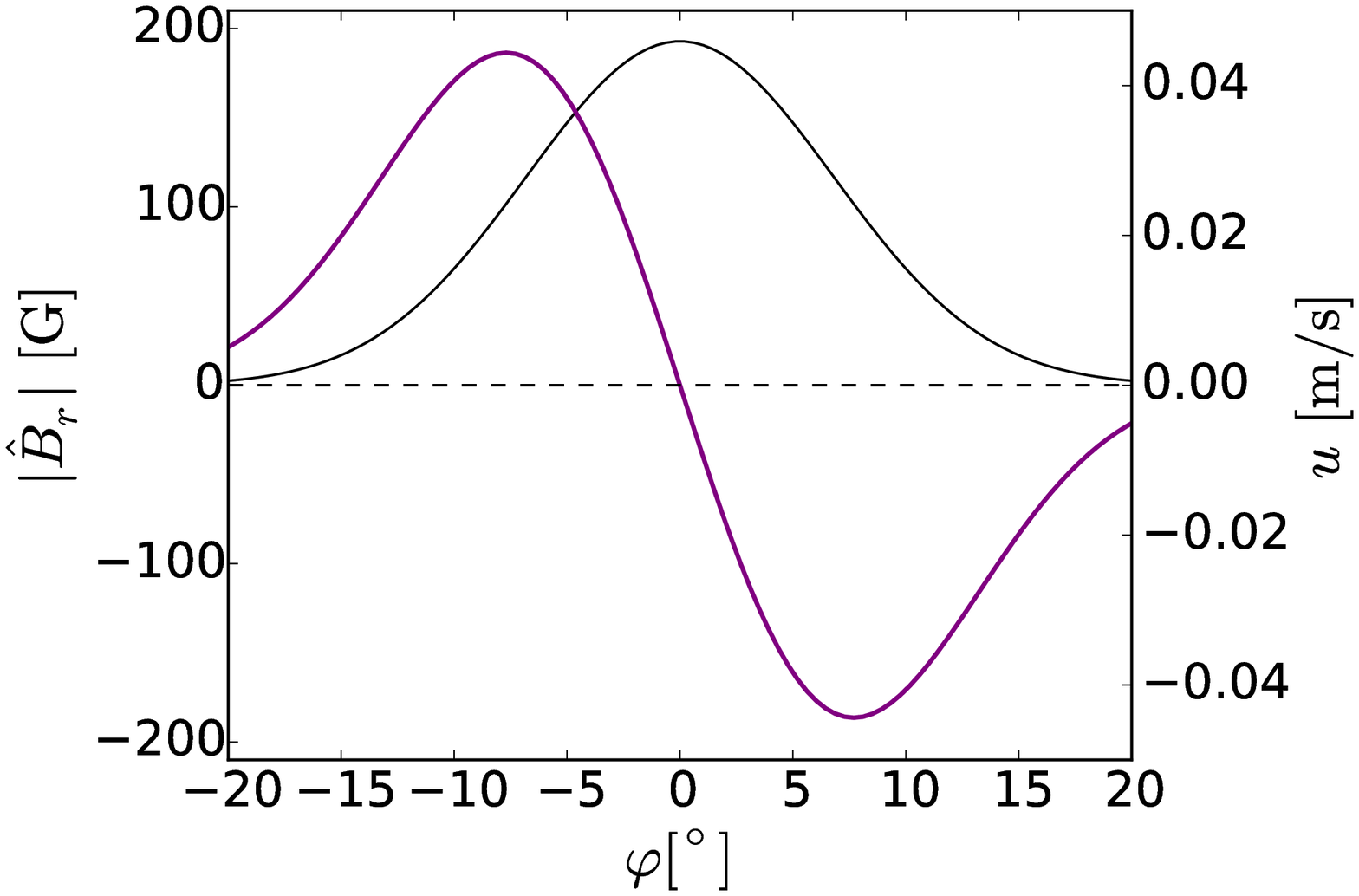}
  }
  \caption{\emph{Left:} Two concentrations of flux of different polarity (thick, vertical lines), smoothed absolute value of the magnetic field of the separate concentrations (red and blue Gaussian curves), and smoothed absolute value of the total magnetic field. \emph{Right:} Smoothed absolute value of the total magnetic field (black line) and inflow velocity profile resulting from Equation \eqref{eq:infVelocParam} (purple line). We have assumed a flux of $4\times10^{18}\,\mathrm{Mx}$ for the single concentrations.}
  \label{fig:inflowexplan}
\end{figure*}
The nearly dipolar field at the end of an activity cycle represents the poloidal flux from which the toroidal flux of the next cycle is generated \citep{cameron2015crucial}. Thus, a backreaction mechanism that affects the latitudinal separation of the polarities would limit the strength of the polar fields and explain the observed cycle variability. \cite{cameron2012strengths} propose the converging flows toward active regions as a major candidate for this nonlinear cycle modulation. These near-surface, large-scale flows toward sunspot groups and active regions were first observed by \cite{gizon2001probing}. These flows have magnitudes of ${\sim}50\,\mathrm{m/s}$ and can extend up to heliocentric angles of $30^\circ$ around the active region center \citep[see, e.g.,][]{haber2002evolving, gizon2008observation, gonzalez2010meridional, gizon2010local}. \cite{derosa2006consequences} incorporated such inflows in their SFT model \citep[see][]{schrijver2001simulations} to study their effect on the evolution of an active region. In their model they explicitly imposed a reduced diffusivity in regions of strong magnetic field in addition to the inflows. They found that, even for comparatively weak inflows, the inflows cause the magnetic concentrations to clump together and generate patterns that are inconsistent with observations. \cite{jiang2010effect} included a magnetic-field-dependent, axisymmetric latitudinal perturbation of the meridional flow compatible with the inflows in SFT simulations and found that this reduced the latitudinal separation of the polarities of the emerged BMRs, thus limiting the buildup of magnetic field at the polar caps. \cite{cameron2012strengths} argue that this effect dominates in strong cycles, while in weak cycles the perturbation of the meridional flow enhances cross-equatorial flux transport and accelerates the reversal of the polar fields. The second question we investigate in this paper is the way inflows affect the evolution of a decaying BMR and their impact on the amount of poloidal flux generated.

\section{Surface flux transport model}

In our model, a magnetically active region is composed of a number $N$ of point-like flux concentrations representing the radial photospheric magnetic field. We approximate the local solar surface as a plane domain centered at a latitude $\lambda_0$ and an arbitrary longitude that we take as $0$. The flux concentrations are subject to advection by differential rotation and convective flows, the latter of which we model as a two-dimensional random walk. If two concentrations of opposite polarity approach each other within a distance of $1\,\mathrm{Mm}$, the pair cancels and we remove them from the simulation. We assume the differential rotation profile determined by \cite{snodgrass1983magnetic}:
\begin{equation}
\omega(\lambda) = 13.38 - 2.30 \sin^2 \lambda - 1.62\sin^4 \lambda \,\, [^{\circ}/\text{day}].
\label{eq:difRot}
\end{equation}

In this reference frame a solid body rotation translates into a uniform velocity field. 

In one of the experiments described later, we include a meridional flow. The exact form of this flow is not critical, since we are performing a local study. We use the form from previous studies such as \cite{vanBallegooijen1998fchannels} or \cite{baumann2004parameter}:
\begin{equation}
 v_\lambda(\lambda) =  v_m \sin(2.4\lambda),
\end{equation}
where $v_m=11\mathrm{m/s}$.

\cite{derosa2006consequences} parametrize the inflows toward active regions in the following way:
\begin{equation}
\mathbf{u}_{\text{in}}=a \nabla |\hat{B_r}|^b,
\label{eq:infVelocParam}
\end{equation}
where $|\hat{B_r}|$ is the absolute value of the magnetic flux density, smoothed with a Gaussian having a full width at half maximum of $15^\circ$. In Fig. \ref{fig:inflowexplan} we sketch the inflow profile resulting from Equation \eqref{eq:infVelocParam} in the case of two close flux concentrations of opposite polarity placed at the same latitude. Figure \ref{fig:infexp1} shows the smoothed absolute value of the magnetic fields of the two concentrations and the total smoothed unsigned magnetic field. Figure \ref{fig:infexp2} shows the inflow velocity profile, proportional to the gradient of the smoothed absolute value of the magnetic field. In our simulations we use a FWHM of $15.5^\circ$, and set $b=1$. In this case, we can equivalently calculate the inflow velocity field as the sum of the inflows driven by the single concentrations:
\begin{equation}
\mathbf{u}_{\text{in}}(\varphi,\lambda)=\sum_{i=1}^N \mathbf{u}_{\text{single}}(\varphi-\varphi_i,\lambda-\lambda_i),
\label{eq:infProfConvol}
\end{equation}
where $\mathbf{u}_{\text{single}}(\varphi,\lambda)$ is the single-concentration inflow profile and $(\varphi_i,\lambda_i)$ is the position of the $i-$th concentration. The parameter $a$ is chosen such that the initial aggregate inflow velocity averaged over the $10\%$ of the domain area with highest inflow velocities amounts to ${\sim}50 \mathrm{m/s}$. We stress that we are not considering that the driving of such an extended inflow by a single concentration is a physical process that actually occurs on the Sun. Our aim is to reproduce a field of converging flows toward the BMR that is somewhat similar to what is observed.

\section{Recovering the diffusion limit}

\subsection{Preliminary discussion}

When describing the dispersive effect of convective flows on the magnetic concentrations as the Fickian diffusion of a continuous quantity (here the radial magnetic field), at least two assumptions are made. The first one is that each of the magnetic field concentrations performs an independent random walk, uncorrelated with the motion of all the other concentrations. The second one is that the random walk steps are small compared to the scale of interest. We now examine these assumptions.

We consider a specific pattern of convective cells (a \emph{realization}): The magnetic field elements are advected toward the border of the cells, adopting a network-like arrangement. In a different realization, a given magnetic element travels a different distance, in a different direction, and takes a different time to reach the border of the cell. Repeating this experiment over a large number of realizations, the statistically expected distribution of magnetic elements at a certain time can be inferred. Similarly fragmentation and merging of flux concentrations strongly affect the correlations in particular realizations, but do not affect the averaged distribution. When the length of the random walk steps is small compared to the scale of interest, the evolution of the expectation value of the flux distribution is approximated well by the diffusion of a continuous flux density. In the following sections we employ our SFT model to investigate whether random steps of the size of the convective cells can be considered small enough to describe the dispersal of magnetic flux on intermediate and large scales as a diffusion process.

\subsection{Methods}

In the limit of small random walk steps, the evolution of the expectation value of the magnetic flux distribution can be described as the diffusion of a continuous quantity representing the radial field with a diffusion coefficient $\eta$ given by
\begin{equation}
\eta = \frac{1}{4}\frac{(\Delta l)^2}{\Delta t}
\label{eq:diffCoef}
\end{equation}
\citep{leighton1964transport}. In what follows we consider the number density of concentrations. Solving the diffusion equation shows that an initial Gaussian density distribution $\rho(r,t_0)$ remains Gaussian at all times, and its standard deviation is given by
\begin{equation}
\sigma(t) = \sqrt{2\eta t + \sigma_0^2},
\label{eq:sigmat}
\end{equation}
where $\sigma_0$ is the standard deviation at $t=0$.

In our experiment, we set up a patch of $N = 8000$ concentrations randomly placed about the center of the domain according to a Gaussian density distribution with an initial standard deviation of $\sigma_0 = 20\,\mathrm{Mm}$. The concentrations undergo random walks for $35\,\mathrm{days}$. The experiment was carried out for two different random walk step sizes ($\Delta l = 500\,\mathrm{km}$ and $\Delta l = 20\,\mathrm{Mm}$), corresponding to small, short-lived granules and large, long-lived supergranules. To compare with the diffusion approximation, we consider random walks corresponding to a fixed diffusion coefficient of $\eta=250\,\mathrm{km^2/s}$. This value is similar to the $\sim257\,\mathrm{km^2/s}$ value reported by \cite{jafarzadeh2014migration} from observations. It is also in the range of diffusivities found in radiative MHD simulations by \cite{cameron2011decay}, and compatible with the evolution of the large-scale fields \citep{cameron2010surface}. The time interval between random walk steps is different for granules and supergranules and corresponds to different lifetimes. Equation \ref{eq:diffCoef} gives a lifetime of $\Delta t = 250\,\mathrm{s} \sim 4 \,\mathrm{min}$ for the granule and $\Delta t = 4\cdot 10^5\,\mathrm{s}\sim 4.5 \,\mathrm{days}$ for the supergranule.

We now consider an annulus centered on the origin of coordinates of the domain. The mean flux density $\bar{\rho}_a$ in the annulus is calculated as the number of concentrations enclosed within it divided by its surface area. If the annulus is sufficiently narrow, its mean density can be directly compared with the diffusion prediction $\rho(\bar{r},t)$, where $\bar{r}$ is an arbitrary point within the annulus. If the random walk steps are short enough, the two quantities should be similar, provided that there are enough realizations or, equivalently, the number of random walkers is very large. To better approach this limit, we average over $1000$ realizations of the experiment.

\subsection{Results}

\begin{figure}[t!]
  \centering
  \resizebox{\hsize}{!}{\includegraphics{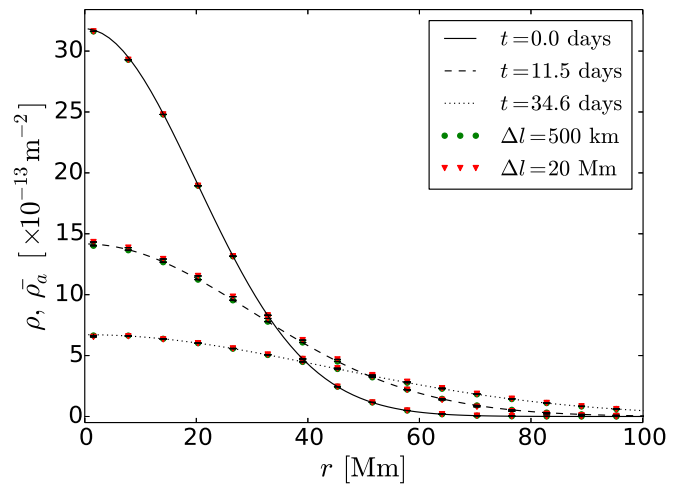}}
  \caption{Comparison between the density distribution resulting from the diffusion equation (line plots) and the averaged number of concentrations per unit area in the simulations. The annuli used to count the concentrations are $2.5\,\mathrm{Mm}$ wide. Only one third of the points have been plotted for better visualization.}
  \label{fig:spread1}
\end{figure} 
\begin{figure}[t!]
  \centering
  \resizebox{\hsize}{!}{\includegraphics{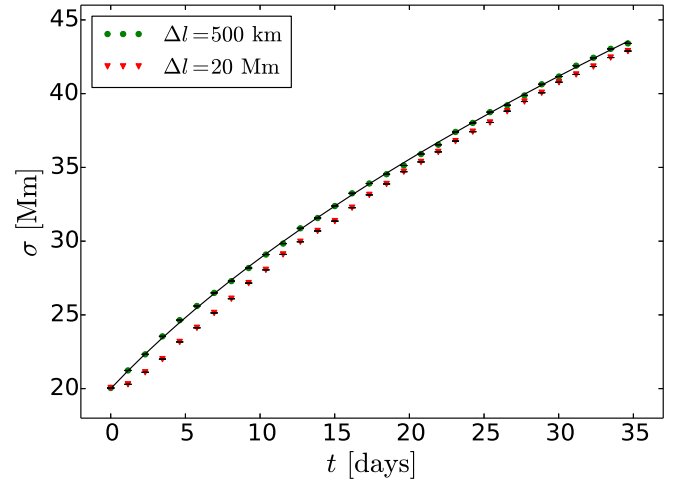}}
  \caption{Full width at half maximum of the averaged distribution of concentrations as a function of time (symbols) compared to the spread of the equivalent diffusion process (continuous line).}
  \label{fig:spread2}
\end{figure}

Figure \ref{fig:spread1} shows the solution of the diffusion equation and the averaged number of flux concentrations per unit area at times $t=0$, $t=11.5\,\mathrm{days}$, and $t=34.6\,\mathrm{days}$. The agreement is very good, and the point concentrations continue matching a Gaussian distribution over time. Fitting the data points from the simulations to Gaussian curves, we can compare the spread of the distributions with the diffusion prediction \eqref{eq:sigmat}. In Fig. \ref{fig:spread2} the standard deviation of the distributions is plotted as a function of time. We see that in the $\Delta l = 500\,\mathrm{km}$ case, the spread matches the diffusion approximation fairly well, whereas in the $\Delta l = 20\,\mathrm{Mm}$ case it is slightly lower than expected from a diffusion process. This is because at this scale the random walk step cannot be considered small with respect to the characteristic scale of the BMR. We also note that the larger discrepancy occurs over the first days of evolution (when the size of the patch is closer to the size of the random walk steps), while we normally are interested in substantially longer evolution times when using SFT models. Moreover, the random walk does not seem to diverge from the diffusion solution. Therefore, we conclude that the diffusion approximation can be safely used when studying the mid- and long-term evolution of magnetic field distributions on the length scales of a typical active region (tens of megameters) or larger, and it is the appropriate treatment when we are interested in, e.g., the evolution of the polar field.

\section{Evolution of a bipolar magnetic region}

\subsection{Setup}

\begin{figure}[!t]
\begin{center}
\resizebox{\hsize}{!}{\includegraphics{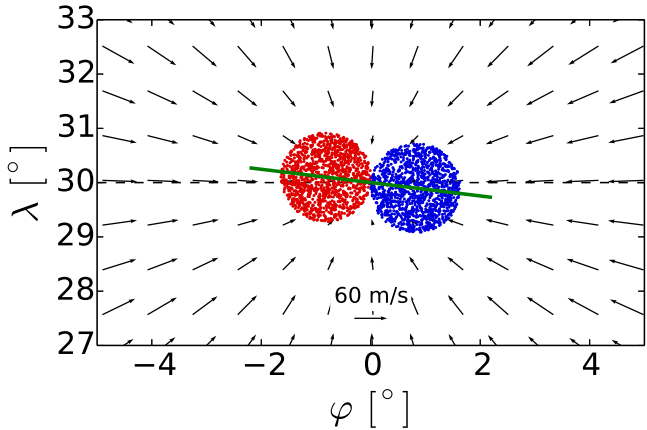}}
\caption{Initial distribution of magnetic flux concentrations in one of the realizations of the experiment with $\alpha_0 = 7^\circ$ and $\lambda_0 = 30^\circ$. The red and blue colors represent opposite polarities. The gray dashed line indicates the latitude of emergence. The solid green line indicates an angle of $7^\circ$ with respect to the longitudinal direction and approximately bisects the BMR. The quiver plot represents the inflows.}
\label{fig:distribution}
\end{center}
\end{figure}

We now consider the effects of the inflows on an isolated BMR. Figure \ref{fig:distribution} shows the initial configuration of our model of BMR for one of the experiments we carried out. The flux concentrations are evenly distributed in two circular regions of opposite polarities, each of which has a radius of $10\,\mathrm{Mm}$ and contains $1250$ concentrations. The BMR is placed at latitude $\lambda_0$ with an initial tilt angle $\alpha_0$. To study the evolution of the BMR, we track the changes of the longitudinal and latitudinal separations of the centers of gravity of the two polarity patches, as well as the tilt angle, the total unsigned flux, and the axial dipole moment. We ran $500$ realizations of each experiment in order to reduce statistical noise. The tilt angle is calculated as the angle between the negative half of the $\varphi$-axis and the line connecting the two centers of gravity (the \emph{dipole axis}); i.e.,
\begin{equation}
\alpha = \arctan\left(\frac{\bar{\lambda}^+-\bar{\lambda}^-}{\bar{\varphi}^--\bar{\varphi}^+}\right),
\end{equation}
where
\begin{equation}
\bar{\varphi}^\pm = \frac{1}{N^\pm}\sum_{i=1}^{N^\pm}\varphi_i^\pm;\,\,\,\,\bar{\lambda}^\pm = \frac{1}{N^\pm}\sum_{i=1}^{N^\pm}\lambda_i^\pm.
\end{equation}
Here, $\varphi_i^\pm$ and $\lambda_i^\pm$ are the coordinates of the $i$-th concentration of the polarity indicated by the superscript, and $N^\pm$ the total number of concentrations of each polarity.

The contribution of the bipolar region to the cancellation and build up of the polar fields depends on its total flux and the latitudinal separation of the polarity patches. To estimate this contribution we calculate the BMR's axial dipole moment, defined as
\begin{equation}
B_p = \int_0^{2\pi}\int_0^\pi B_r(\varphi, \theta) Y_1^0 \sin\theta \,\mathrm{d}\theta\mathrm{d}\varphi,
\label{eq:admCont}
\end{equation}
where $\theta$ is the colatitude, $\theta = \pi/2-\lambda$. In our discrete representation and in terms of the latitude $\lambda$, the integral becomes
\begin{equation}
B_p = \frac{\phi_0}{R_\odot^2}\sqrt{\frac{3}{4\pi}}\sum_{i=1}^N p_i \sin(\lambda_i)
\label{eq:admDisc}
\end{equation}
where $\phi_0$ is the flux of one concentration, and $p_i$ the polarity ($\pm 1$) of the $i-$th concentration. Considering typical values for BMRs with moderate-to-strong magnetic fields, we assume a total unsigned flux of $\Phi_0 = 10^{22}\,\mathrm{Mx}$ \citep[see, e.g.,][]{schrijver_activity}, which gives a single concentration magnetic flux or $\phi_0 = 4\cdot10^{18}\,\mathrm{Mx}$. Under diffusion alone, the Sun's axial dipole moment would decay on a time scale $\tau_d = \frac{R_\odot^2}{\eta}$ \citep{leighton1964transport}. For a diffusion coefficient $\eta = 250\,\mathrm{km^2/s}$, $\tau_d\sim30\,\mathrm{years}$. In our plane domain approximation, the solar radius is infinite, so $\tau_d$ is infinite as well. Since $Y_1^0$ does not depend upon longitude, the axial dipole moment is expected to be conserved in the simulations where only differential rotation and random walks are included. In a more realistic spherical geometry, the axial dipole would decline on a time scale that is large compared to the length ($35\,\mathrm{days}$) of our simulations.

\subsection{Results}

% Effect of the inflows
% ---------------------

\subsubsection*{Flux dispersal}

\begin{figure}[!tb]
\begin{center}
\resizebox{\hsize}{!}{\includegraphics{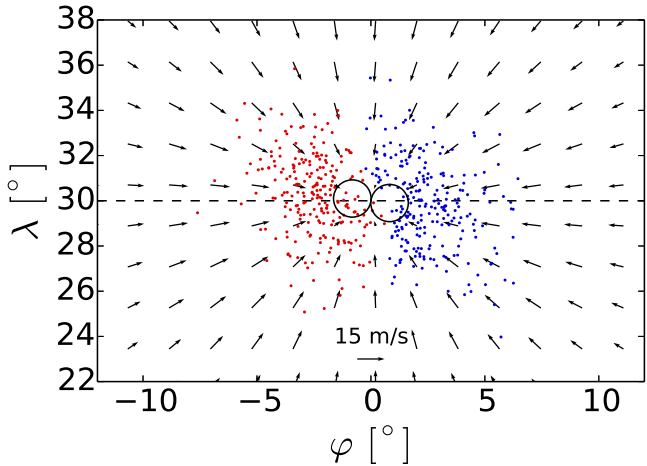}}
\caption{Positions and polarities of the magnetic flux concentrations shown in Fig. \ref{fig:distribution} after evolving for $17\,\mathrm{days}$ with inflows turned on. The two circles outline the initial configuration of the BMR as shown in Fig. \ref{fig:distribution}. From the initial $2500$ concentrations, only $480$ remain.}
\label{fig:evol}
\end{center}
\end{figure}

\begin{figure}[t]
\begin{center}
\resizebox{\hsize}{!}{\includegraphics{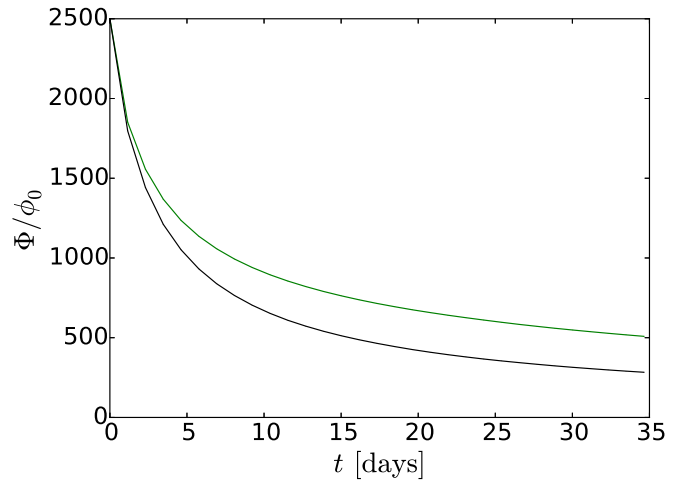}}
\caption{Total unsigned flux of a BMR placed at $\lambda_0 = 30^\circ$ with an initial tilt of $\alpha_0 = 7^\circ$. The black and green lines correspond to the simulations with and without inflows, respectively. The error bars indicating realization noise are too small to be visible. The enhanced flux destruction when inflows are present seen in the plot is very similar in all the cases studied.}
\label{fig:fluxDestruction}
\end{center}
\end{figure}

The question we want to answer is how the inflows affect the dispersal of the magnetic field. In Fig. \ref{fig:evol} we see the distribution of the magnetic concentrations starting from that shown in Fig. \ref{fig:distribution} after evolving for $17\,\mathrm{days}$ under the influence of differential rotation, random walk, and inflows. The BMR has been sheared by the differential rotation, while the random walk has dispersed the flux concentrations. At $t=17\,\mathrm{days}$ and in spite of the inflows, the BMR spans close to $10^\circ$ in the longitudinal direction and around $8^\circ$ in the latitudinal direction. The total flux decreases very rapidly owing to cancellation (see Fig. \ref{fig:fluxDestruction}), especially in the first days, and the inflow velocity decays accordingly.

\begin{figure}[!t]
\begin{center}
\resizebox{\hsize}{!}{\includegraphics{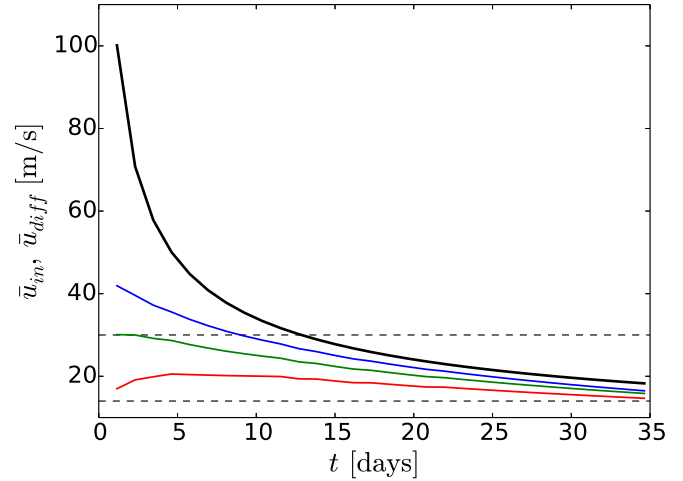}}
\caption{\textit{Thick black line:} $\bar{u}_{d}(t)$ for the case under discussion ($\alpha_0=7^\circ$, $\lambda_0=30^\circ$). \textit{Colored lines:} $\bar{u}_{in}(t)$ for $0$ (red), $10$ (green) and $20\,\mathrm{Mm}$ from the origin. \textit{Dashed lines:} Shear flow velocities at $\sim12^\circ$ (lower) and $\sim14^\circ$ polewards from the central latitude.}
\label{fig:meanflows}
\end{center}
\end{figure}

To compare the dispersive effect of the convective flows against the inflows, we begin by considering the two separately. At time $t$, a concentration undergoing a random walk alone is separated from its initial position by an average distance given by the diffusion length, $l_d(t)=2\sqrt{\eta t}$. We define a mean expansion velocity $\bar{u}_{d}(t) = l_d(t)/t = 2\sqrt{\eta/t}$. The space- and time- averaged inflow velocity along such a path over a time $t$ is given by
\begin{equation}
\bar{u}_{in}(t) = \frac{1}{t\,l_d(t)}\int_{l_0}^{l_0+l_d(t)}\int_0^t \mathbf{u}_{\text{in}}[\varphi(l'),\lambda(l'),t']\cdot \mathrm{d}\mathbf{l'}\mathrm{d}t',
\end{equation}
where $l'$ denotes the distance from the starting position $l_0$. To evaluate this integral, we take the inflow velocity $\mathbf{u}_{\text{in}}[\varphi(l),\lambda(l),t]$ from one realization of the experiment. We choose the path $l_d$ along the $\varphi$ axis for simplicity and perform the calculation for $l_0=0$, $10\,\mathrm{Mm}$, and $20\,\mathrm{Mm}$ from the origin. Both $\bar{u}_{d}(t)$ and $\bar{u}_{in}(t)$ are plotted in Fig. \ref{fig:meanflows}.

The mean expansion velocity decays as $t^{-1/2}$, but it remains higher than the mean inflow velocity at all times and for different values of $l_0$, most prominently during the first days. The differential rotation also contributes to the escape of magnetic flux. The shear flow velocity at $\lambda=32\,^\circ$ is ${\sim}14\mathrm{m/s}$, and it reaches ${\sim}30\,\mathrm{m/s}$ at $\lambda=34^\circ$. It is seen that the turbulent dispersal and the differential rotation dominate the (decaying) inflows. We therefore do not observe the clumping reported by \cite{derosa2006consequences}. A possible cause for this discrepancy is the explicit additional damping of dispersal of large magnetic field concentrations that these authors include in their model \citep[see][]{schrijver2001simulations} and we do not. This feature seeks to reproduce the reduced flux dispersal observed in areas of large magnetic field. \citep[In][the authors report a flux dispersal characterized by a diffusion coefficient of ${\sim}250\,\mathrm{km^2/s}$ in areas surrounding the core of an active region, while within the core region the diffusion coefficient is ${\sim}110\,\mathrm{km^2/s}$]{schrijver1990patterns}.

We point out here that the inflows alone will have the effect of reducing the apparent diffusivity in active regions. We can estimate this effect by considering the velocities involved. Assuming a random walk step of $15\,\mathrm{Mm}$ and a diffusion coefficient of $250\,\mathrm{km^2/s}$, equation \eqref{eq:diffCoef} gives a travel time of ${\sim}2.6\,\mathrm{days}$. The concentrations travel with a velocity of ${\sim}67\,\mathrm{m/s}$. A random walk with this travel time and characterized by a diffusion coefficient of $110\,\mathrm{km^2/s}$ has a step size of ${\sim}10\,\mathrm{Mm}$ and a travel velocity of ${\sim}44\,\mathrm{m/s}$. The difference between travel velocities in the two cases is $\sim23\,\mathrm{m/s}$, a value comparable with the averaged magnitude of the inflows.

% Dependence upon initial tilt
% ----------------------------
\subsubsection*{Dependence upon initial tilt}

\begin{figure*}[p]
  \centering
  \subfloat[$\alpha_0 = 0^\circ$\label{fig:alpha1}]{%
    \includegraphics[width=0.9\textwidth]{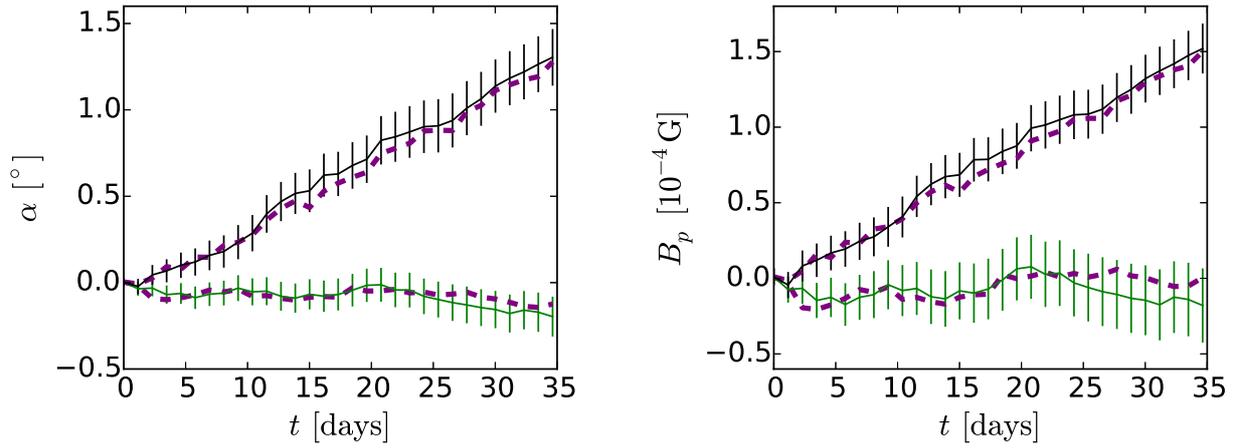}
  }\\
  \subfloat[$\alpha_0 = 7^\circ$\label{fig:alpha2}]{%
    \includegraphics[width=0.9\textwidth]{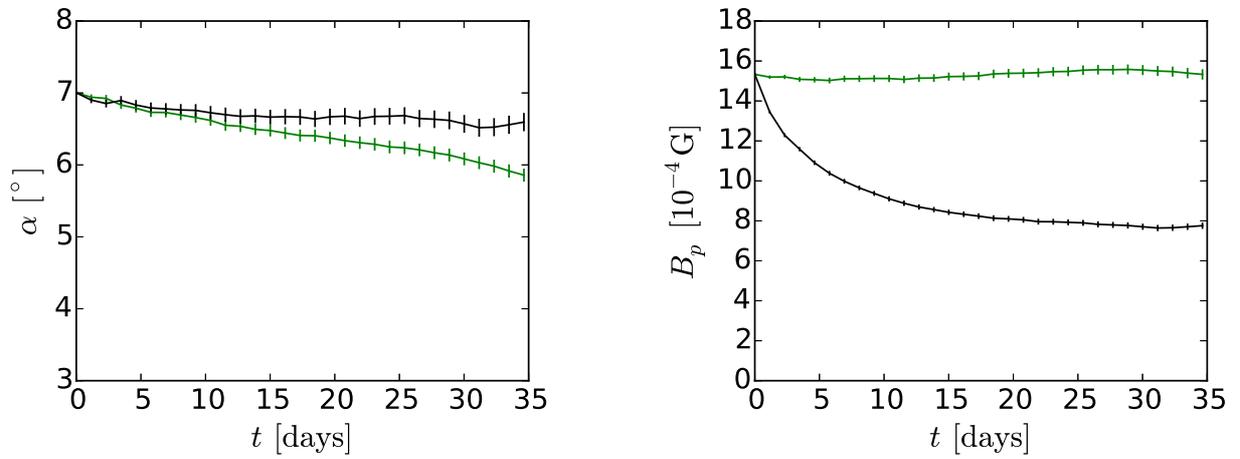}
  }\\
  \subfloat[$\alpha_0 = -7^\circ$\label{fig:alpha3}]{%
    \includegraphics[width=0.9\textwidth]{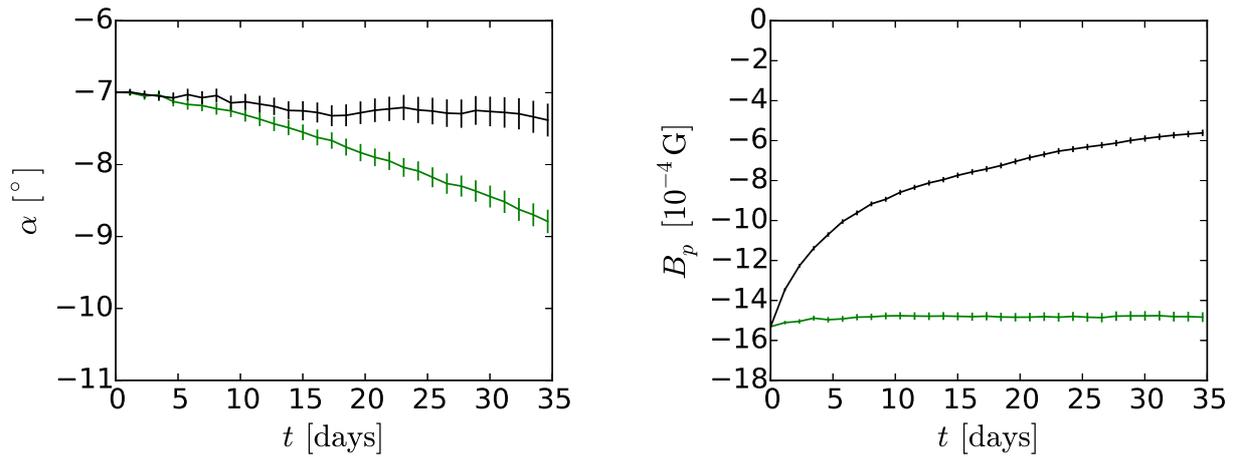}
  }
  \caption{Tilt angle and axial dipole moment of a BMR placed at $\lambda_0 = 30^\circ$ with an initial tilt of $\alpha_0 = 0^\circ$ \protect\subref{fig:alpha1}, $\alpha_0 = 7^\circ$ \protect\subref{fig:alpha2}, $\alpha_0 = -7^\circ$ \protect\subref{fig:alpha3}. The black and green lines correspond to the simulations with and without inflows respectively. The purple dashed lines in \protect\subref{fig:alpha1} correspond to simulations where the meridional flow is included (error bars have not been plotted for clarity). The error bars represent the standard error of the mean values and indicate the realization noise. The errors for different times are correlated.}
  \label{fig:alpha}
\end{figure*} 

Figure \ref{fig:alpha} shows the tilt angle of the BMR and the axial dipole moment as given by equation \eqref{eq:admDisc} for the case of a bipolar region placed at $\lambda_0=30^\circ$ and initial tilt angles of $\alpha_0 = 0^\circ$ (Fig. \ref{fig:alpha1}), $\alpha_0 = 7^\circ$ (Fig. \ref{fig:alpha2}), and $\alpha_0 = -7^\circ$ (Fig. \ref{fig:alpha3}). We start by discussing the $\alpha_0 = 7^\circ$ case. The longitudinal separation is initially greater than the separation in latitude, and the shear flow causes it to increase faster than the latter. This causes the dipole axis to rotate counter-clockwise, i.e., to decrease in tilt angle. When inflows are switched on, the growth of the longitudinal and latitudinal separations is restricted in such a way that the rotation of the dipole axis is slower, and flux cancellation is enhanced (see Fig. \ref{fig:fluxDestruction}). The latter is expected since the converging flows tend to bring concentrations closer together, increasing the probability of pair cancellation.

As expected, the axial dipole moment remains constant when inflows are not present. When inflows are present $B_p$ decays until reaching an approximately constant value after ${\sim}20\text{-}30\,\mathrm{days}$, when the both the inflows-on and inflows-off plots essentially become parallel. At this point the inflows are so weak that the subsequent evolution of the BMR is dominated by the differential rotation and the random walk. It should be noted that inflows restrict the growth of the latitudinal separation of polarities over time rather than reducing it \citep[see][]{jiang2010effect}. The decrease in $B_p$ is a consequence of how the latitudinal separation and the enhanced flux destruction balance. In the case under discussion, the inflows cause a decrease in axial dipole moment.

The tilt angle and axial dipole moment for the case of an initial tilt angle of $-7^\circ$ are presented in Fig. \ref{fig:alpha3}. Now the trailing polarity is placed at a lower latitude than the preceding patch, so the shear flow tends to make the angle increase towards more negative values. When the inflows are included, the reduced latitudinal separation causes the tilt angle to be less negative than in the case without inflows, i.e., the dipole axis rotates more slowly. As before, flux destruction is enhanced by inflows, and the absolute value of the axial dipole moment is reduced.

\begin{figure*}[ht]
  \centering
  \subfloat[\label{fig:afig1}]{%
    \frame{\includegraphics[width=5cm]{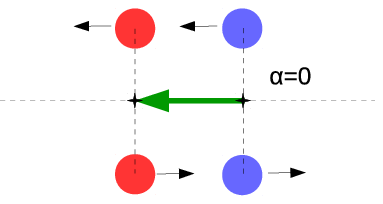}}
  }\hspace{0.66cm}
  \subfloat[\label{fig:afig2}]{%
    \frame{\includegraphics[width=5cm]{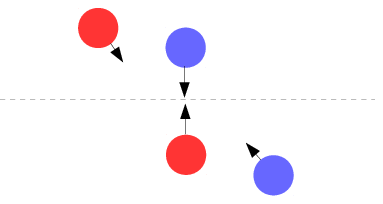}}
  }\hspace{0.66cm}
  \subfloat[\label{fig:afig3}]{%
    \frame{\includegraphics[width=5cm]{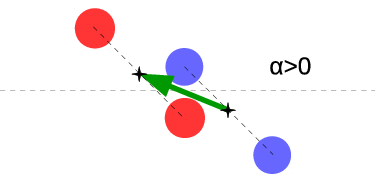}}
%  }\hspace{0.33cm}
%  \subfloat[\label{fig:afig4}]{%
%    \frame{\includegraphics[width=4cm]{alpha3.eps}}
  }
  \caption{Schematic representation of the combined action of differential rotation and inflows leading to the latitudinal separation of the polarity patches giving rise to a non-zero axial dipole moment. The blue and red dots represent flux concentrations. The thin black arrows indicate plasma flows. The black plus signs indicate the location of the centers of gravity of each polarity. The thick green arrow outlines the dipole axis. \protect\subref{fig:afig1}. The flux concentrations of the BMR emerge at different latitudes. \protect\subref{fig:afig2} Differential rotation shears the structure, and inflows tend to bring magnetic concentrations closer together. The effect of the inflows on the outermost concentrations is weaker. \protect\subref{fig:afig3} This displaces the centroid of the polarity patches away from the central latitude, causing the dipole axis to rotate.}
  \label{fig:alpha_effect}
\end{figure*}

Figure \ref{fig:alpha1} shows the same magnitudes for the $0^\circ$ initial tilt angle. The longitudinal separation is again restricted by the inflows, but the latitudinal separation increases with time, which produces a clockwise rotation of the dipole axis. This behavior departs from what could be expected from either diffusion, differential rotation or converging flows acting in isolation. In Fig. \ref{fig:alpha_effect} we provide a schematic explanation of this effect. The differential rotation shears the BMR, and at the same time the inflows tend to bring the innermost concentrations closer to the central latitude, while the outermost concentrations are less affected. This displaces the centers of gravity of the polarity patches away from the central latitude: the trailing polarity is shifted poleward and the preceding polarity moves towards the equator. The rotation in the case considered here amounts to ${\sim}1.3^\circ$ over $35$ days. When inflows are off, the axial dipole moment remains close to zero, as expected from a BMR with no initial tilt. When inflows are present, however, there is an increase in latitudinal separation, and consequently $B_p$ also increases.

Using observed positions and tilt angles of active regions as input to an SFT model, \cite{cameron2010surface} were able to reproduce the main features of the open flux inferred from the \emph{aa}-index of geomagnetic variations during solar cycles 15 to 21. However, the authors need to scale the tilt angles by a factor of $0.7$ to reduce the amount of flux arriving at the poles and so match the observed amplitude of the open flux. The righthand column of Fig. \ref{fig:alpha} shows that the axial dipole moment of the BMR is indeed substantially decreased as a consequence of the inflows (except in the $\alpha_0 = 0$ case). The problem of how inflows affect the axial dipole moment of complex active regions and sunspot groups, rather than in an isolated BMR, has yet to be studied. Nevertheless, it is seen that the inflows can provide the physical mechanism needed to justify this scaling of the tilt angles.

Assuming the BMR emerges away from the equator, its contribution to the total axial dipole moment of the Sun is proportional to $\sin\theta$ and, under advection by the meridional flow alone, declines on a time scale $\tau_f = R_\odot/v_m\approx2\,\mathrm{years}$ \citep{wang1991axial}. This characteristic time becomes infinite in our plain domain approximation. As a result, the meridional flow does not have an appreciable effect on the axial dipole moment during the first month of evolution of the BMR, when inflows are non-negligible. This is shown in Fig. \ref{fig:alpha1}, where the purple dashed lines represent the evolution of the corresponding quantities in simulations including meridional flow. These do not show any appreciable difference with the plots obtained in the simulations without meridional flow.

% Dependence upon initial latitude
% --------------------------------

\subsubsection*{Dependence upon latitude of emergence}
\begin{figure*}[p]
  \centering
  \subfloat[$\lambda_0 = 0^\circ$\label{fig:lambda1}]{%
    \includegraphics[width=0.9\textwidth]{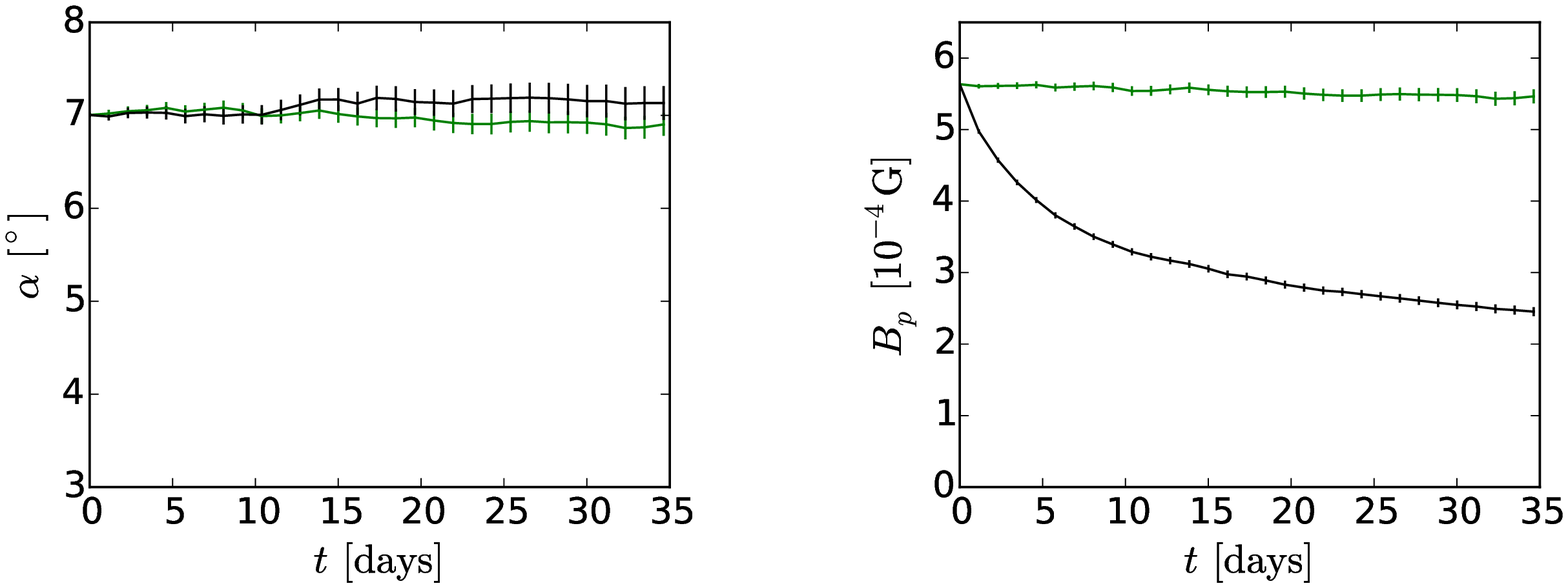}
  }\\
  \subfloat[$\lambda_0 = 15^\circ$\label{fig:lambda2}]{%
    \includegraphics[width=0.9\textwidth]{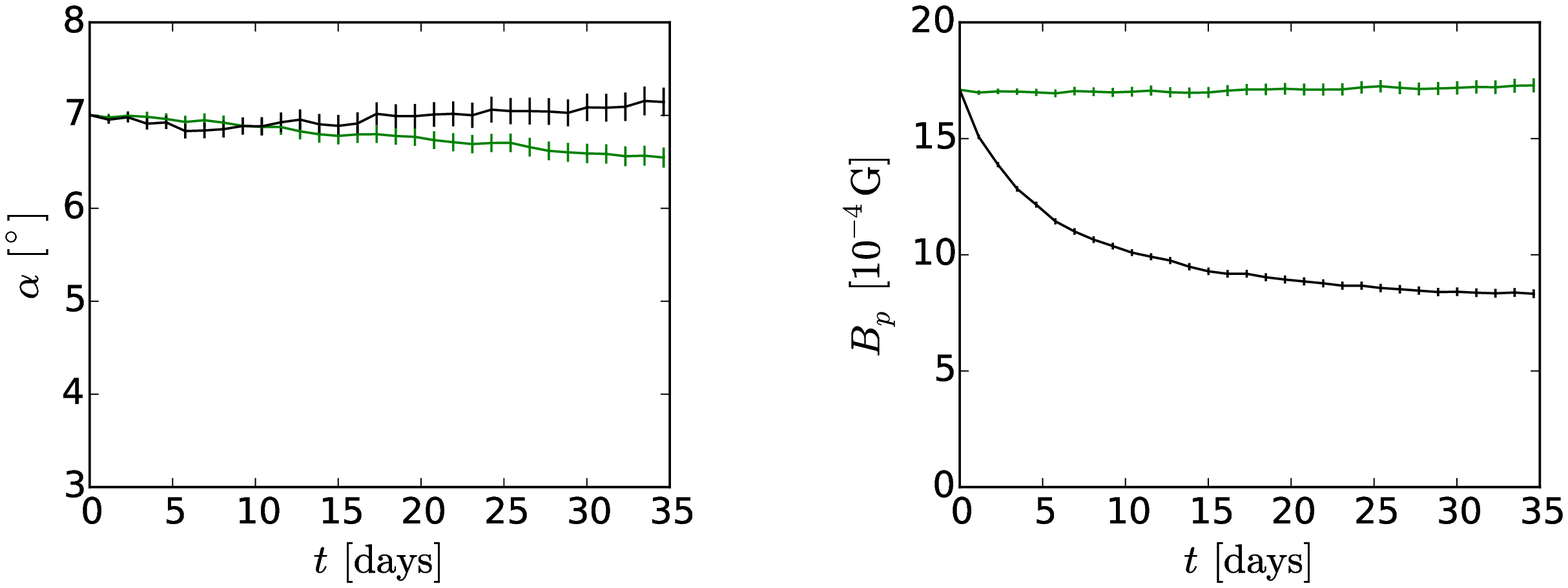}
  }\\
  \subfloat[$\lambda_0 = 45^\circ$\label{fig:lambda3}]{%
    \includegraphics[width=0.9\textwidth]{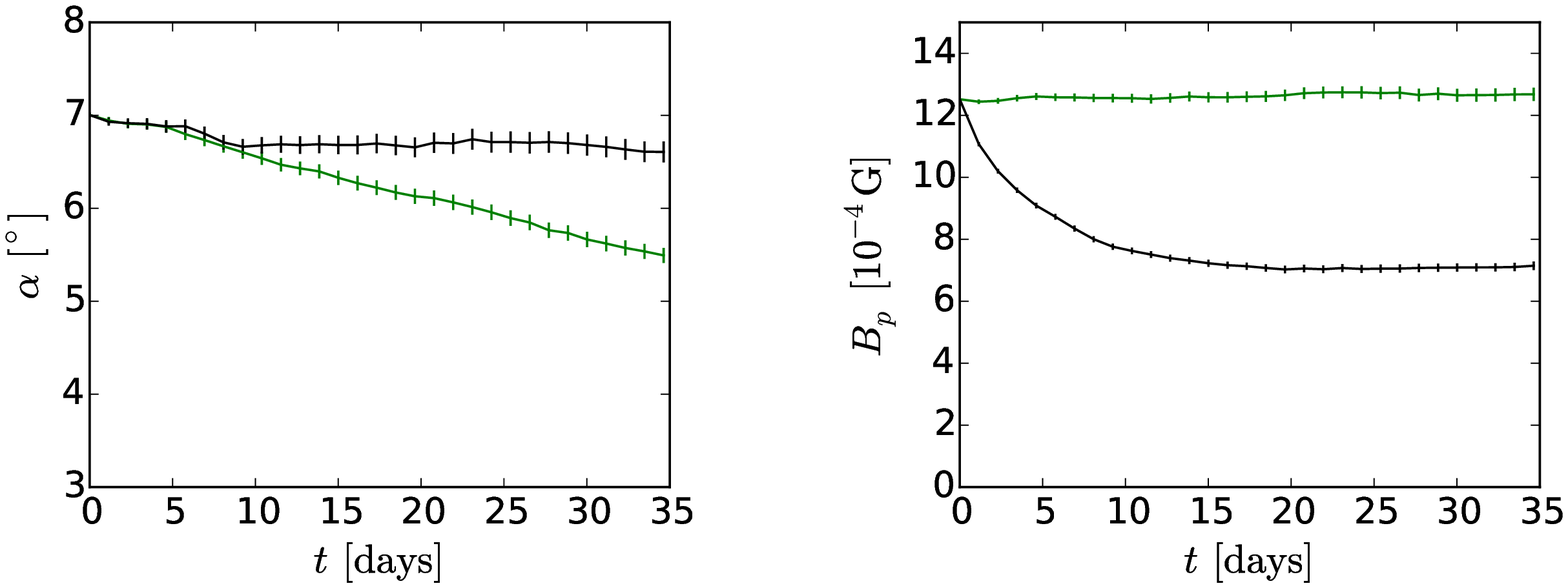}
  }
  \caption{Tilt angle, total unsigned flux, and poloidal flux for a BMR with an initial tilt angle of $7^\circ$ placed at $\lambda_0 = 0^\circ$ \protect\subref{fig:lambda1}, $\lambda_0 = 15^\circ$ \protect\subref{fig:lambda2}, and $\lambda_0 = 45^\circ$ \protect\subref{fig:lambda3}. The black and green lines correspond to the simulations with and without inflows, respectively. The error bars represent the standard error of the mean values and indicate the realization noise. The errors for different times are correlated.}
  \label{fig:lambda}
\end{figure*} 
In Fig. \ref{fig:lambda} we present the same magnitudes as before for a BMR with an initial tilt angle of $7^\circ$ placed at $\lambda_0=0$ (\ref{fig:lambda1}), $\lambda_0=15^\circ$ (\ref{fig:lambda2}) and $\lambda_0=45^\circ$ (\ref{fig:lambda3}). The case $\lambda_0=30^\circ$ is presented in Fig. \ref{fig:alpha2}. In all four cases, the inflows restrict the latitudinal and longitudinal separations of the centroids. This results in the tilt angle remaining roughly constant for the BMR at the equator, in the cases both with and without inflows. For the BMRs placed at higher latitudes, the tilt angle decreases over time; i.e., the dipole axis rotates counter-clockwise, and the rotation is slower when the inflows are on. As before, when inflows are switched on, the total unsigned flux decreases faster and the axial dipole moment decays over time. The stronger shearing at higher latitudes causes the inflows to decay faster, so the decrease in $B_p$ is not as pronounced as at lower latitudes. Nevertheless, this decrease is very similar in all four cases, so the dependence of the effect of inflows upon the latitude of the BMR is not very strong.

\section{Conclusion}

In this study we used an SFT code to test the suitability of the diffusion treatment to describe the dispersal of magnetic flux by convective flows on the solar surface. The dispersal produced by the random walk is slightly lower than expected from a Fickian diffusion process when the step size is comparable to the typical size of an active region. However this discrepancy is not very important, even for lengths corresponding to the biggest and longest-lived supergranules. We therefore conclude that the diffusion treatment is appropriate for describing the flux dispersal due to supergranulation when the scales of interest resemble the size of a typical active region or higher. 

We also investigated the role of the converging flows toward a BMR in its evolution and their impact on the axial dipolar field. We saw that the differential rotation and the dispersion by convective motions suffice to oppose the inflows, which decay very quickly owing to flux cancellation. We pointed out that the inflows may cause the apparent reduced rate at which strong magnetic fields appear to diffuse. We have also seen that, in most cases, the inflows limit the latitudinal separation of the polarities and enhance flux cancellation, which lowers the contribution of the emerged BMR to the axial dipole moment. This is an important effect for generating the polar fields. However, when the initial tilt angle is close to zero, the inflows increase the latitudinal separation of the polarities, which gives rise to a positive tilt angle and generates an axial dipole moment. Finally, it was shown that meridional flow does not have an appreciable influence on the early stages of the evolution of an emerged BMR, when the inflows are not negligible.

\section*{Aknowledgements}
We want to thank Manfred Schüssler for his valuable suggestions and his thorough revisions of this manuscript. This work was carried out in the context of Deutsche Forschungsgemeinschaft SFB 963 "Astrophysical Flow Instabilities and Turbulence" (Project A16).

\bibliographystyle{aa}
\bibliography{/home/david/texmf/bibtex/reviews,/home/david/texmf/bibtex/papers,/home/david/texmf/bibtex/theses,/home/david/texmf/bibtex/books}

\begin{thebibliography}{28}
\expandafter\ifx\csname natexlab\endcsname\relax\def\natexlab#1{#1}\fi

\bibitem[{{Baumann} {et~al.}(2004){Baumann}, {Schmitt}, {Sch{\"u}ssler}, \&
  {Solanki}}]{baumann2004parameter}
{Baumann}, I., {Schmitt}, D., {Sch{\"u}ssler}, M., \& {Solanki}, S.~K. 2004,
  \aap, 426, 1075

\bibitem[{Cameron {et~al.}(2010)Cameron, Jiang, Schmitt, \&
  Sch{\"u}ssler}]{cameron2010surface}
Cameron, R., Jiang, J., Schmitt, D., \& Sch{\"u}ssler, M. 2010, The
  Astrophysical Journal, 719, 264

\bibitem[{Cameron \& Sch{\"u}ssler(2012)}]{cameron2012strengths}
Cameron, R. \& Sch{\"u}ssler, M. 2012, Astronomy \& Astrophysics/Astronomie et
  Astrophysique, 548

\bibitem[{{Cameron} \& {Sch{\"u}ssler}(2015)}]{cameron2015crucial}
{Cameron}, R. \& {Sch{\"u}ssler}, M. 2015, Science, 347, 1333

\bibitem[{{Cameron} {et~al.}(2011){Cameron}, {V{\"o}gler}, \&
  {Sch{\"u}ssler}}]{cameron2011decay}
{Cameron}, R., {V{\"o}gler}, A., \& {Sch{\"u}ssler}, M. 2011, \aap, 533, A86

\bibitem[{Charbonneau(2010)}]{charbonneau2010lrsp}
Charbonneau, P. 2010, Living Reviews in Solar Physics, 7

\bibitem[{De~Rosa \& Schrijver(2006)}]{derosa2006consequences}
De~Rosa, M. \& Schrijver, C. 2006, in Proceedings of SOHO 18/GONG 2006/HELAS I,
  Beyond the spherical Sun, Vol. 624, 12

\bibitem[{{DeVore} {et~al.}(1984){DeVore}, {Boris}, \&
  {Sheeley}}]{devore1984meridional}
{DeVore}, C.~R., {Boris}, J.~P., \& {Sheeley}, Jr., N.~R. 1984, \solphys, 92, 1

\bibitem[{Fan(2009)}]{fan2009lrsp}
Fan, Y. 2009, Living Reviews in Solar Physics, 6

\bibitem[{{Gizon} {et~al.}(2010){Gizon}, {Birch}, \& {Spruit}}]{gizon2010local}
{Gizon}, L., {Birch}, A.~C., \& {Spruit}, H.~C. 2010, Annual Review of
  Astronomy and Astrophysics, 48, 289

\bibitem[{{Gizon} {et~al.}(2001){Gizon}, {Duvall}, \&
  {Larsen}}]{gizon2001probing}
{Gizon}, L., {Duvall}, Jr., T.~L., \& {Larsen}, R.~M. 2001, in IAU Symposium,
  Vol. 203, Recent Insights into the Physics of the Sun and Heliosphere:
  Highlights from SOHO and Other Space Missions, ed. P.~{Brekke}, B.~{Fleck},
  \& J.~B. {Gurman}, 189

\bibitem[{Gizon \& Rempel(2008)}]{gizon2008observation}
Gizon, L. \& Rempel, M. 2008, Solar Physics, 251, 241

\bibitem[{{Gonz{\'a}lez Hern{\'a}ndez} {et~al.}(2010){Gonz{\'a}lez
  Hern{\'a}ndez}, {Howe}, {Komm}, \& {Hill}}]{gonzalez2010meridional}
{Gonz{\'a}lez Hern{\'a}ndez}, I., {Howe}, R., {Komm}, R., \& {Hill}, F. 2010,
  The Astrophysical Journal Letters, 713, L16

\bibitem[{{Haber} {et~al.}(2002){Haber}, {Hindman}, {Toomre}, {Bogart},
  {Larsen}, \& {Hill}}]{haber2002evolving}
{Haber}, D.~A., {Hindman}, B.~W., {Toomre}, J., {et~al.} 2002, The
  Astrophysical Journal, 570, 855

\bibitem[{{Jafarzadeh} {et~al.}(2014){Jafarzadeh}, {Cameron}, {Solanki},
  {Pietarila}, {Feller}, {Lagg}, \& {Gandorfer}}]{jafarzadeh2014migration}
{Jafarzadeh}, S., {Cameron}, R.~H., {Solanki}, S.~K., {et~al.} 2014, \aap, 563,
  A101

\bibitem[{{Jiang} {et~al.}(2010){Jiang}, {I{\c s}ik}, {Cameron}, {Schmitt}, \&
  {Sch{\"u}ssler}}]{jiang2010effect}
{Jiang}, J., {I{\c s}ik}, E., {Cameron}, R.~H., {Schmitt}, D., \&
  {Sch{\"u}ssler}, M. 2010, The Astrophysical Journal, 717, 597

\bibitem[{{Leighton}(1964)}]{leighton1964transport}
{Leighton}, R.~B. 1964, \apj, 140, 1547

\bibitem[{{Mackay} {et~al.}(2002{\natexlab{a}}){Mackay}, {Priest}, \&
  {Lockwood}}]{mackay2002single}
{Mackay}, D.~H., {Priest}, E.~R., \& {Lockwood}, M. 2002{\natexlab{a}},
  \solphys, 207, 291

\bibitem[{{Mackay} {et~al.}(2002{\natexlab{b}}){Mackay}, {Priest}, \&
  {Lockwood}}]{mackay2002full}
{Mackay}, D.~H., {Priest}, E.~R., \& {Lockwood}, M. 2002{\natexlab{b}},
  \solphys, 209, 287

\bibitem[{{Mart{\'{\i}}nez Pillet} {et~al.}(1997){Mart{\'{\i}}nez Pillet},
  {Lites}, \& {Skumanich}}]{mpillet1997active}
{Mart{\'{\i}}nez Pillet}, V., {Lites}, B.~W., \& {Skumanich}, A. 1997, The
  Astrophysical Journal, 474, 810

\bibitem[{Schrijver(2001)}]{schrijver2001simulations}
Schrijver, C.~J. 2001, The Astrophysical Journal, 547, 475

\bibitem[{{Schrijver} \& {Martin}(1990)}]{schrijver1990patterns}
{Schrijver}, C.~J. \& {Martin}, S.~F. 1990, \solphys, 129, 95

\bibitem[{{Schrijver} \& {Zwaan}(2008)}]{schrijver_activity}
{Schrijver}, C.~J. \& {Zwaan}, C. 2008, {Solar and Stellar Magnetic Activity}

\bibitem[{Snodgrass(1983)}]{snodgrass1983magnetic}
Snodgrass, H.~B. 1983, The Astrophysical Journal, 270, 288

\bibitem[{{Solanki}(1993)}]{solanki1993smallscale}
{Solanki}, S.~K. 1993, \ssr, 63, 1

\bibitem[{{van Ballegooijen} {et~al.}(1998){van Ballegooijen}, {Cartledge}, \&
  {Priest}}]{vanBallegooijen1998fchannels}
{van Ballegooijen}, A.~A., {Cartledge}, N.~P., \& {Priest}, E.~R. 1998, \apj,
  501, 866

\bibitem[{{Wang} {et~al.}(1989){Wang}, {Nash}, \&
  {Sheeley}}]{wang1989evolution}
{Wang}, Y.-M., {Nash}, A.~G., \& {Sheeley}, Jr., N.~R. 1989, \apj, 347, 529

\bibitem[{{Wang} \& {Sheeley}(1991)}]{wang1991axial}
{Wang}, Y.-M. \& {Sheeley}, Jr., N.~R. 1991, \apj, 375, 761

\end{thebibliography}

\end{document}